\documentstyle[aaspp4]{article}
\lefthead{KALOGERA}
\righthead{FORMATION OF LMXBS}
\begin{document}
\begin{center}
To appear in {\em The Astrophysical Journal}
\end{center}
\vspace{1.cm}
\title{Formation of Low-Mass X-ray Binaries. III. A New Formation Mechanism: 
Direct Supernova} 
\author{Vassiliki Kalogera}
\affil{Astronomy Department, University of Illinois at Urbana-Champaign, \\
1002 West Green St., Urbana, IL 61801. \\
e-mail: vicky@astro.uiuc.edu}
\newcommand{\df}{distribution function}
\newcommand{\dfs}{distribution functions}
\newcommand{\am}{angular momentum}
\newcommand{\os}{orbital separation}
\newcommand{\oss}{orbital separations}
\newcommand{\br}{birth rate}
\newcommand{\Mo}{M$_\odot$}
\newcommand{\Ro}{R$_\odot$}
\newcommand{\ns}{neutron star}
\begin{abstract}
We propose a new formation mechanism (direct-supernova) 
for low-mass X-ray binaries (LMXBs) 
that does 
not involve any prior phase of mass transfer. Survival through the 
supernova (SN) explosion and shrinkage of the orbit is achieved by a kick 
velocity 
of appropriate magnitude and direction imparted to the neutron star at its 
birth. We present analytical population synthesis calculations of 
LMXBs forming via both 
the direct-SN and the helium-star SN mechanisms, and compare the results. We 
find that the direct-SN channel contributes a non-negligible fraction of the 
total LMXB population, depending strongly on the r.m.s. magnitude of the kick 
velocity. More importantly, the direct-SN mechanism provides a natural way for 
the formation of low-mass binary pulsars in nearly circular orbits with orbital 
periods in excess of $\sim 100^{\mbox{d}}$, which cannot 
have been formed via the helium-star SN mechanism. 
\end{abstract} 
\section{INTRODUCTION}

Since their discovery low-mass X-ray binaries (LMXBs) 
have been a puzzle for theories of close
binary evolution. The  
existence of a low-mass star in a small orbit around a compact object (\ns\ or
black hole) appeared to 
require a quite intriguing explanation concerning the evolutionary path 
followed by the progenitors  of these systems. 
The present orbits of LMXBs are too small to have accommodated the 
growth in size of the progenitors of the compact objects. In addition, the 
masses of the companions to the compact objects (donor stars) are so small 
that the survival probability through a 
supernova (SN) explosion is expected to be small. 
The small 
number of observed LMXBs ($\sim 100$, van Paradijs\markcite{P95} 1995) 
along with their 
relatively long lifetimes suggests that the evolutionary path responsible for 
their formation is a quite improbable one (Webbink\markcite{W92} 1992).  

Over the years, three formation mechanisms have been put forward in an 
effort to understand the existence of LMXBs. All of them invoke evolution 
through a common envelope (CE) phase (Paczy$\acute{\mbox{n}}$ski\markcite{P76} 
1976), 
during which the low-mass  
star spirals inward through the extended envelope 
of the massive primary star, and 
the phase is terminated upon ejection of the common envelope. This phase 
results in the reduction of both  
the mass of the progenitor of the compact object and the orbital 
separation. One of the three formation mechanisms 
involves 
the collapse into a \ns\ of a massive white dwarf, accreting mass from a 
low-mass companion, in a small orbit (the outcome of an earlier CE 
phase). The collapse of a white dwarf into a neutron star 
in the context of formation of X-ray binaries was first proposed by Flannery \& 
van den Heuvel\markcite{F75} (1975) and Canal \& Schatzman\markcite{C76} 
(1976).  
A second mechanism, the He-star SN, involves a CE phase 
leaving a helium star in a small orbit with a low-mass companion. The helium 
star evolves to core collapse and undergoes a supernova explosion, forming 
a neutron star remnant (Sutantyo 1975\markcite{S75}; 
van den Heuvel\markcite{H83} 1983). 
The third evolutionary sequence invokes the formation of a 
Thorne-$\dot{\mbox{Z}}$ytkow 
object as the end product of a massive X-ray binary with a  
third component in a very wide orbit. This low-mass third star is engulfed 
in the envelope of the Thorne-$\dot{\mbox{Z}}$ytkow object. The ejection of the 
common envelope leaves the low-mass star in orbit with the \ns\ 
(Eggleton \& Verbunt\markcite{E83} 1986; however, see Fryer, Benz, \& 
Herant\markcite{F96} 1996).  

Recent reassessments of pulsar kinematics have reinforced and magnified 
earlier suggestions that neutron stars are endowed at birth with large kick 
velocities, the apparent result of asymmetric core collapse. 
Studies of the radio pulsar population (e.g., 
Harrison, Lyne, \& Anderson\markcite{H93} 1993; Lyne \& Lorimer\markcite{L94} 
1994) show that pulsars 
have space velocities much higher than those of their massive progenitors and 
that they extend to large distances away from the Galactic plane\footnote{
Recently, Iben \& Tutukov\markcite{I95} (1995) have argued against the 
existence of kick velocities. However, they need to assume a binary fraction 
equal to unity, and even then their results are marginally consistent 
with early estimates of pulsar velocities (Harrison et al. 1993).}. 
Using a new electron 
density model, they conclude from pulsar proper motions that the mean kick 
velocity is $\sim 450\pm 90$\,km/s, a result that appears to be corroborated 
by studies of pulsar-supernova remnant associations 
(Frail\markcite{F96} 1996; although see Hartman\markcite{H96} 1996; 
Bhattacharya\markcite{B96} 1996).  
Moreover, results of simulations of supernova 
explosions (e.g., Herant, Benz, \& Colgate\markcite{HB92} 1992; 
Janka \& M\"{u}ller\markcite{J94} 1994; 
Burrows, Hayes, \& Fryxell\markcite{B95} 1995;
Burrows \& Hayes\markcite{BH96} 1996) 
also support the idea that kick velocities are imparted to neutron 
stars at birth, although more detailed numerical calculations are 
needed to settle the issue.  
 
In this paper we explore the possibility that a simple evolutionary sequence 
can lead to the formation of LMXBs. The essence of the mechanism lies in 
the possibility  that even 
if the orbits of the primordial binaries are so wide that the two stars do 
not interact and a common envelope is not formed, the systems remain bound 
and the orbital separations decrease after the supernova explosion due to a 
kick of appropriate magnitude and direction imparted to the neutron star at 
birth. 

The proposed evolutionary path is described in detail in the next section. The 
constraints and limits on the parameter space of the progenitors specific to 
this mechanism  
are identified in \S\,3. We have performed population synthesis calculations,  
the method and results of which are presented in \S\,4. A discussion of the 
implications and our conclusions are presented in \S\,5. 
 
\section{FORMATION MECHANISM}

Let us consider a primordial binary with an extreme mass ratio, in which the 
primary is massive
enough to explode as a supernova and form a 
neutron star at the end of its evolution, and the secondary is a
low-mass star ($ M_2 \lesssim 2$\Mo ).

During its evolution, the primary loses mass in a stellar wind and the orbital 
separation of the binary increases. If the initial orbit is wide enough,  
the primary never fills its Roche lobe, despite
its growth in radius, and its evolution is terminated when it reaches the core 
collapse stage. Thus, the binary components do not interact in any way prior 
to the supernova explosion, except perhaps for some small (in our case, 
negligible because of the wide orbits considered) 
accretion by the secondary from the wind of the primary.

The mass loss during the collapse of the primary is so severe that the system 
would be disrupted if the explosion were symmetric.  
However, in the presence of a kick velocity imparted to the 
newborn \ns , there is a finite probability that the post-SN system remains 
bound. The  
survival probability depends primarily on the magnitude and direction of the 
kick velocity and less on the amount of mass lost.

Although the supernova explosion is the most crucial event in the evolution of 
an LMXB progenitor, keeping the post-SN system bound is not enough for an 
observable LMXB to be formed. The orbit after the explosion must be small 
enough so that the low mass star can fill its Roche lobe (i) 
in a time shorter than 
the Galactic disk age and (ii) before it reaches the end of its evolution and 
acquires its maximum radius.  
A kick velocity of the appropriate magnitude and direction can not 
only keep the post-SN system from becoming unbound, but can also decrease the 
orbital separation.   
The subsequent decrease of the post-SN \os\ due to a combination of (i) tidal 
dissipation and orbital circularization, and (ii) \am\ losses (caused by 
gravitational radiation and the magnetic stellar wind of the 
secondary), aided by nuclear evolution of the secondary,  
eventually brings the system into contact. For the first time in the 
evolutionary history of the binary, the stellar components interact and the 
system may appear as a luminous X-ray source, depending on the characteristics 
of the mass transfer phase. We name this formation mechanism 
{\em direct supernova}, since the binary members do not experience any phase of 
interaction prior to the supernova explosion.  

\section{CONSTRAINTS ON THE PARAMETER SPACE OF THE PROGENITORS}

A primordial binary follows the evolutionary path described above and becomes
a LMXB only if it satisfies a number of constraints. The simplicity of the 
formation mechanism results in a set of simple constraints, as well. 

There is
only one constraint imposed on the characteristics of the binaries before the 
supernova explosion. 
The initial \os\ of the system 
must be large enough so that the primary does not fill its Roche lobe before 
it reaches core collapse. Otherwise, unstable mass transfer is initiated and
the system will evolve according to the He-star SN mechanism.

We have used the evolutionary calculations presented by Schaller et 
al.\markcite{S92} (1992) for solar composition  
to fit the maximum radius acquired by a massive star undergoing
wind mass 
loss, as a function of its 
initial mass (see Appendix in Kalogera \& Webbink\markcite{KW96b} 1996b; 
hereafter Paper II). Using this relation and the  
radius of the Roche lobe of the primary expressed in units of 
the orbital separation (Eggleton\markcite{E83} 1983),
we can calculate the orbital separation of systems with their primaries just 
filling their Roche lobes at the time of their maximum extent. This separation 
represents a lower limit to the orbital size of those primordial binaries 
that will evolve according to the direct-SN mechanism. The limiting orbital 
separations for $1$\,\Mo\ and $2$\,\Mo\ companions and for a range of primary
masses are shown in Figure 1. It is evident that the LMXB progenitors specific 
to the direct-SN mechanism have initial orbital separations and periods 
in excess of 
$\sim 600 - 2000$\,R$_\odot$ and $\sim 1-5$\,yr, respectively.  

The fact that the pre-SN binary orbits are so wide, 
along with the large amounts 
of mass lost at supernovae, results into highly eccentric orbits 
immediately after the explosions. These orbits are similar to those of 
tidal capture binaries formed in dense stellar environments, as globular 
clusters. Recent detailed studies of the tidal capture process presented by 
Mardling\markcite{M95} (1995a,b) show that there is a region in the parameter 
space of 
eccentricity, $e$, and ratio of the periastron distance to the stellar radius, 
$R_p/r_\ast$, where binaries exhibit chaotic behavior, with large changes in 
eccentricity, that may even lead to self-ionization. Binaries outside this 
region circularize only via dissipation of energy. During this long quiescent 
phase, the eccentricity varies quasiperiodically due to a quasiperiodic 
exchange of energy between the orbit and the tides, and a merger is avoided. To 
secure that the post-SN binaries formed via the direct-SN mechanism survive 
and eventually become circularized, their post-SN characteristics must be such 
that they populate the non-chaotic region of the $R_p/r_\ast -e$ parameter space. Indeed, we find that although the mean eccentricity of the binaries produced 
by our synthesis models is high, $\langle e\rangle =0.93$, the ratio 
of the periastron distance to the stellar radius also  
acquires high values,  $\langle R_p/r_\ast \rangle \simeq 100$. 
These values greatly exceed the limit on $R_p/r_\ast$, below which chaotic 
behavior is possible  
($R_p/r_\ast< 5$; Mardling 1995a). Tidal 
circularization of these binaries proceeds initially at a time scale much 
longer than the Hubble time, 
but by the time Roche-lobe filling occurs and mass transfer is initiated, the 
time scale has become extremely small and the orbits are most probably 
circular. 
 
The post-SN binaries at the onset of the mass-transfer phase must satisfy 
a set of structural and evolutionary 
constraints, which are independent of the specific LMXB formation mechanism and 
must be satisfied by neutron star-normal star binaries if they are to appear 
as LMXBs. These constraints have been studied by Kalogera \& 
Webbink\markcite{KW96} (1996a; 
hereafter Paper I) 
and they concern (a) the age of the systems, which must not exceed the age of 
the Galactic disk, and (b) the ability of the donors to remain in 
hydrostatic and thermal equilibrium at the onset of mass transfer. Systems are 
further divided into two groups, those transferring mass at sub-Eddington 
rates (conservative mode) and those with donors driving mass transfer at 
super-Eddington rates (non-conservative mode). We note that the process of 
super-Eddington accretion is not well understood and it is possible that 
matter surrounding super-Eddington systems may quench the X-rays and that these 
systems do not appear as LMXBs.  

\section{POPULATION SYNTHESIS}

\subsection{The Model}

We have performed population synthesis calculations for LMXBs forming 
according to the direct supernova mechanism, using the analytic 
method presented in 
Paper II. 
We transform the distribution 
of primordial binaries through the various evolutionary stages, 
i.e.,
wind mass-loss from
the primary, supernova explosion of the primary imparting a kick velocity to
the newborn neutron star, shrinkage of the orbit due to angular momentum
losses, and nuclear evolution of the low-mass companion until
Roche-lobe filling by the companion is achieved and the nascent LMXB is formed.
In order to model the physical processes involved we have employed analytic
approximations of results from evolutionary calculations, which are given in
Papers I and II. We have assumed a Maxwellian distribution for the kick 
velocities;  
the method for incorporating their effects  
developed by Kalogera\markcite{K96} (1996) has been used. 
We have made the same assumptions for the parent binary population as in the 
study of LMXB formation via the helium-star SN mechanism (Paper II), except
for one modification appropriate to the specifics of the direct-SN mechanism. 
These assumptions have been extensively discussed in Paper II. Here, we 
only summarize the key points and describe the modification applied. 

We assume that primordial binaries are characterized by three parameters: the 
primary mass, $M_1$, the mass ratio, $q\equiv M_2/M_1$, where $M_2$ is the 
mass of the secondary, and the \os\ $A$. It is conceivable that a fourth 
parameter is the eccentricity of the orbits.  
For the wide systems of interest 
to us, the time scale for circularization (e.g., Zahn\markcite{Z77} 1977, 1989)
is initially much longer than the lifetime of the primary. However, 
we find that the direct-SN channel is primarily fed by binaries with orbital 
separations comparable (within a factor of less than two) 
to the limiting values for 
Roche lobe overflow at the time of maximum extent of the primary. For these
binaries, 
as the 
massive star evolves to the giant branch the circularization time scale becomes 
shorter than about one hundredth of its main sequence lifetime, which is short 
enough for the orbits become circular prior to the supernova event. 
Moreover, for a scale-less distribution in orbital separations, as we will 
assume, the distribution of separations is not altered by the 
circularization process. Therefore, we may assume that all LMXB 
progenitors feeding the direct-SN formation channel are formed with 
 circular orbits. 

For the primordial binary population, we adopt the same initial distributions 
as in Paper II (eq.\,[4],\,[5]), except for the integral in the expression 
for the the distribution of binaries over mass ratios and orbital separations, 
$g(q,A)$. The difference arises from the fact that, in the direct-SN 
evolutionary channel, additional companions to the primary in inner stable 
orbits need not to be excluded, as in the case of the He-SN channel. The 
presence of such companions and their possible interaction with the primary 
does not affect the evolution of the binary under study, which follows the 
direct-SN channel as long as its orbit is wide enough.  
In fact, it is conceivable that when the neutron 
star forms there is more than one companion for it to remain in a bound orbit 
with, but we will not consider the evolution of multiple systems here.
The distribution of primary binaries over mass ratio, $q$, and orbital 
separation, $A$, is then given by:
\begin{equation}
g(q,A)~=~\frac{0.075}{A}~0.04q^{-2.7}~\exp \left( - 
\int_{A\cdot (6.3)^{-2/3}}^{A\cdot (6.3)^
{2/3}}\int_{q}^{1}~0.075~A'^{-1}0.04~q'^{-2.7}~dA'dq' \right).
\end{equation}
The exponential term (Poisson probability) excludes from the distribution any 
companions more massive than the secondary in dynamically unstable orbits 
(see also Paper II). A plot of both the above distribution and the one 
appropriate for the helium-star SN mechanism, for specified primary mass and 
orbital separation, is shown in Figure 2. Since the ``exclusion zone'' 
in orbital separation is narrower for the direct-SN channel the frequency 
of available progenitors is increased. 

\subsection{Results}

The analytical method of our synthesis computations enables us to calculate 
the two-dimensional distribution, $\Phi_P(\log M_d, \log P_X)$, 
of neutron star-normal star binaries over 
donor masses, $M_d$, and 
orbital periods, $P_X$, at the onset of the mass 
transfer phase. The distribution of systems with donors in hydrostatic and 
thermal equilibrium, initiating mass transfer in less than $10^{10}$\,yr and 
transferring mass at both sub- and super-Eddington rates is shown in Figure 3a. 
We have chosen an intermediate value of the r.m.s. kick velocity, 
$<V_k^2>^{1/2}$, equal to $300$\,km/s. In Figure 3b the distribution of 
the corresponding group of binaries having formed 
via the helium-star SN mechanism is 
also shown (taken from Paper II), for $<V_k^2>^{1/2}=300$\,km/s and 
$\alpha_{CE}=1$, where $\alpha_{CE}$ is the common envelope efficiency.  

The qualitative characteristics of both distributions in Figures 3a and 3b  
bear many similarities, which 
are primarily dictated by physical processes, such 
as angular momentum losses and nuclear evolution of the low mass star,  
common to both formation mechanisms. As we have also discussed in 
Paper II, the evolution of short period binaries is dominated by angular 
momentum losses due to a magnetic stellar wind from the donor and they populate
a narrow range of orbital periods forming a prominent ``ridge'' along the zero 
age main sequence. As the orbital period increases magnetic braking becomes 
less effective and a ``valley'' is created at $P_X\sim 1^{\mbox{d}}$. At longer 
periods, it is the expansion of the donors due to nuclear evolution that is 
responsible for Roche-lobe overflow. These systems with evolved donors populate 
the ``hump'' in the distributions at long periods 
and masses from $\sim 1$ to $\sim 1.5$\,M$_\odot$. We note that these  
systems initially drive mass transfer 
at super-Eddington rates 
(see Paper I). 

By integrating $\Phi_P(\log M_d,\log P_X)$ over $\log M_d$ we obtain the 
distributions of systems over orbital periods, shown in Figure 4 for both 
the direct-SN and the He-star SN mechanisms. The origin of the peaks at short 
orbital periods is related not only to the effect of magnetic braking but 
also to the flattening of the radius-mass relation along the zero-age main 
sequence. The plateau that appears as soon as magnetic braking becomes 
efficient, between $3^{\mbox{h}}$ to $5^{\mbox{h}}$, is related to the 
increased incidence of primordial binaries with very low-mass companions 
relative to those for the He-star SN formation mechanism (see Figure 2). 
Systems with evolved donors formed via the He-star supernova peak at 
orbital periods of 
$\sim 5^{\mbox{d}}$, whereas for the direct-SN mechanism systems 
with much longer periods are favored. This is the result of the obvious 
difference in orbital 
separation of the pre-SN binaries in the two mechanisms: systems 
following the direct-SN evolutionary channel are much wider 
($A_{pre-SN}^{DSN}\sim 1000$\,R$_\odot$) than those following the He-star SN 
channel ($A_{pre-SN}^{DSN}\sim 10$\,R$_\odot$), which underwent 
dramatic orbital 
shrinkage occurring during the common envelope phase.  

Apart from the comparison of the qualitative characteristics of nascent LMXBs, 
it is also important to compare the results quantitatively based on the birth 
rates of the two evolutionary sequences. Although the absolute birth rates 
strongly depend on the assumptions regarding the primordial binary population, 
and are relatively insensitive to the evolutionary stages involved in each 
channel, the relative birth rates are quite useful in determining the 
efficiency of the mechanisms in LMXB formation. In addition, the fact that 
both mechanisms 
have been modeled under the same set of assumptions renders the comparison 
meaningful. For our typical cases of $<V_k^2>^{1/2}=300$\,km/s and 
$\alpha_{CE}=1$, the birth rates of sub- and super-Eddington systems together 
are $4.5\times 10^{-8}$\,yr$^{-1}$ and $2.5\times 10^{-6}$\,yr$^{-1}$ for the 
direct-SN 
and the He-star SN mechanisms, respectively.  
 
\section{DISCUSSION}

In the direct-SN formation mechanism proposed here, there is only one free 
parameter, besides the assumed parent population, namely 
the r.m.s. kick velocity, 
$<V_k^2>^{1/2}$. We have performed synthesis calculations for a wide range of 
values of $<V_k^2>^{1/2}$ from $10$\,km/s up to $500$\,km/s. The predicted 
birth rates show a strong dependence on the kick velocity (Figure 5); they 
span a range from $\sim 10^{-8}$\,yr$^{-1}$ to $\sim 10^{-6}$\,yr$^{-1}$
for the total population, and from $\sim 10^{-9}$\,yr$^{-1}$
to $\sim 10^{-7}$\,yr$^{-1}$ for systems transferring mass at sub-Eddington 
rates only.
The masses of the neutron star progenitors are such that all the pre-SN 
binaries would be disrupted in the case of a symmetric explosion. However, the 
survival probability through an asymmetric supernova peaks when 
$<V_k^2>^{1/2}$ is comparable to the average relative orbital velocity, 
$<V_r>$, of the stars in the pre-SN binaries (Kalogera 1996). For the 
progenitors specific to the direct-SN mechanism we find that 
$<V_r>\simeq 37$\,km\,s$^{-1}$ and indeed the predicted birth rate peaks at 
$<V_k^2>^{1/2}\simeq 50$\,km/s (see Figure 5). 

We can estimate the efficiency of the direct-SN mechanism relative to that  
involving a He-star SN by comparing the corresponding birth rates. Their 
ratio as a function of $<V_k^2>^{1/2}$ is shown in Figure 6. For  
r.m.s. kick velocities exceeding $\sim 300$\,km/s the 
direct-SN channel appears to be responsible for a few per cent of the 
LMXB population. For  
smaller kick velocities, the direct-SN mechanism contributes a growing share 
of the total population, reaching $25\%$ for 
$<V_k^2>^{1/2}\sim 100$\,km/s for both sub- and super-Eddington systems. 
In the case of the sub-Eddington systems the 
direct-SN channel in fact dominates for $<V_k^2>^{1/2} \lesssim 
100$\,km\,s$^{-1}$
because of the inefficiency of the He-star SN channel in producing short-period 
systems when kick velocities are small (Paper II). 
We note that for less efficient common envelope ejection ($\alpha_{CE}<1$), the 
predicted birth rates for the He-star SN mechanism decreases and hence the 
contribution of the direct-SN channel becomes more significant.  
It is evident that for the current estimates of $<V_k^2>^{1/2}$ the 
direct-SN mechanism 
accounts for a small but  
non-negligible fraction of the total LMXB observed population. 
However, recent studies of the radio pulsar population (Hartman 1996) and of 
the galactic distribution of LMXBs (Bhattacharya 1996) provide evidence that 
the fraction of low velocity pulsars may be higher than that implied by Lyne \& 
Lorimer (1994). 
Such an excess of low kick 
velocities greatly enhances the importance of the direct-SN formation 
mechanism. 

Using our synthesis models, we can calculate the typical orbital parameters 
of the progenitors of LMXBs produced by the evolutionary channel studied here. 
For the primordial binaries, the mean primary and secondary masses 
are $12$\,M$_\odot$ and 
1.2\,M$_\odot$, respectively, and the mean orbital separation is 
1500\,R$_\odot$. 
The mean relative orbital velocity just prior to the supernova explosion 
is 37\,km\,s$^{-1}$. These values, 
along with the two limits imposed on systemic velocities of post-SN binaries 
(Kalogera 1996), result in recoil velocities for these systems in 
the range 20-50\,km\,s$^{-1}$, which are significantly lower than those of 
LMXBs formed via the He-star SN mechanism ($\gtrsim 100$\,km\,s$^{-1}$; 
see Kalogera 1996). These low systemic velocities also indicate that 
LMXBs produced via the direct-SN channel in globular clusters 
can remain bound to the clusters, contrary to the ones produced by the He-star 
SN channel, provided that their wide progenitors survive 
in such a dense stellar 
environment. 

For the synthesis calculations presented here, we have assumed that the 
kick velocities follow a Maxwellian distribution. As discussed above,
survival after the explosion is 
favored only if the kick velocity has magnitude comparable and direction  
opposite to that of the 
relative orbital velocity of the pre-SN system. Comparison between the mean 
orbital parameters of the progenitors and the lower limits imposed on them
(see Figure 1) indicates that the orbital separation of systems that 
eventually become LMXBs is restricted in a very narrow range (factor of 
$\sim 1.5$ from the mean value). Consequently their systemic velocities 
are also concentrated in a narrow range (factor of $\sim 1.2$ from the 
mean value), and therefore kick velocities that favor survival have magnitudes 
between 30 and 45\,km\,s$^{-1}$. This range of velocities is so narrow 
that our results become insensitive to the shape of the velocity distribution
and depend merely on the relative fraction of velocities with magnitude in 
the appropriate range. The observed distribution of pulsar velocities 
(Lyne \& Lorimer 1994) cannot distinguish between a Maxwellian kick 
distribution and a distribution that is flat up to the average velocity and 
and has a smooth cutoff (remark made by an anonymous referee). 
Using a simple estimate of the relative fraction 
of kick velocities within the appropriate range of values, we expect that 
for a flat distribution the birth rate of LMXBs will be increased by a 
factor of $\sim 10$ compared to the birth rate calculated when a 
Maxwellian distribution is assumed with 
$\langle V_k^2 \rangle ^{1/2}=300$\,km\,s$^{-1}$. 
 
Apart from the connection of the new formation mechanism to LMXB production,  
it is also relevant to the formation of low-mass binary pulsars in circular 
orbits, and in particular to those with orbital periods in excess of 
$\sim 100^{\mbox{d}}$ (such as B0820+02, B1800-27, and B1953+29; see van den 
Heuvel\markcite{H95} 1995). 
We have previously pointed out (see Paper II) that these 
long-period systems could not have been formed via the He-star SN mechanism:
an upper limit to the orbital periods of the progenitors is imposed by the 
requirement that 
the primaries fill their Roche lobes prior to the supernova explosion, 
as translated through  
a common envelope phase. The radical 
reduction in binary separation during common envelope evolution 
results in an 
upper limit on the orbital periods of LMXBs with evolved donors of 
$\sim 30^{\mbox{d}}$. Such a low upper limit cannot explain the existence of 
low mass binary pulsars with orbital periods up to $\sim 1200^{\mbox{d}}$,  
even if one takes into account the subsequent evolution of these systems 
through exhaustion of the envelope of the giant donor  
(e.g., Verbunt\markcite{V93} 1993). On the other hand, no upper limit is 
imposed on the 
orbital periods of LMXB progenitors following the direct-SN channel. The 
maximum orbital period of these LMXBs is limited only by the maximum 
possible extent of evolved stars with masses of $\sim 1$\,M$_\odot$ to 
$\sim 1.5$\,M$_\odot$, and therefore orbital periods of up to 
$\sim 600^{\mbox{d}}$ 
can be reached at the onset of mass transfer. 
In the case of $\langle V_k^2\rangle^{1/2}=100$\,km/s, the predicted birth 
rate of LMXBs with orbital periods in excess of $30^{\mbox{d}}$
formed via the direct-SN channel matches the observed 
fraction (2-3/15; van den Heuvel 1995) of the long-period binary 
millisecond pulsars, 
and for higher average kick velocities the birth rate 
lies within a factor of about four from their observed incidence.
Therefore, the secular evolution of 
these long-period LMXBs produced via the direct-SN channel appears to be a 
promising formation mechanism of long-period 
low mass binary pulsars in circular orbits, whose progenitors are absent 
from the LMXB population produced by the He-star SN mechanism.  

\acknowledgments

It is a pleasure to thank Ron Webbink for many stimulating discussions 
and for 
carefully reading this manuscript. I would also like to thank Fred Lamb and 
Dimitrios Psaltis for useful comments. This research was supported by National 
Science Foundation under
grant AST92-18074.

\newpage 

\begin{center}
\large
{\bf Figure Captions}
\normalsize
\end{center}

\figcaption{Minimum orbital separations of primordial binaries that 
follow the direct-SN formation mechanism for two different companion 
masses: $M_2=1$\,M$_\odot$ (solid line) and $M_2=2$\,M$_\odot$ (dotted line).}

\figcaption{Distributions of primordial binaries with primary mass 
$M_1=12$\,M$_\odot$ and orbital separation $A=1500$\,R$_\odot$ over mass 
ratios, $q$. The solid line corresponds to the distribution appropriate for 
the direct-SN channel and the dotted line to that appropriate for the 
helium-star SN channel. The corresponding secondary masses, $M_2$, are also 
shown.} 

\figcaption{Distribution of binaries that transfer mass at both sub- and 
super-Eddington rates over donor masses, $M_d$, and orbital periods, $P_X$, 
for (a) the direct-SN, and (b) the He-star SN ($\alpha_{CE}=1$) 
formation mechanisms.} 

\figcaption{Distribution of binaries transferring mass at both sub- and 
super-Eddington rates over orbital periods, $P_X$, 
for the direct-SN (solid line) 
and the He-star SN (dotted line, $\alpha_{CE}=1$) mechanisms. 
The predicted birth rates are 
$3\times 10^{-8}$\,yr$^{-1}$ and $2.5\times 10^{-6}$\,yr$^{-1}$, respectively.}

\figcaption{Predicted birth rates as a function of r.m.s. kick velocity, 
$<V_k^2>^{1/2}$, for both sub- and super-Eddington systems (filled circles) 
and for sub-Eddington systems only (open circles).} 

\figcaption{Ratio of the direct-SN birth rate to the He-star SN 
(for $\alpha_{CE}=1$) birth rate 
as a function of r.m.s. kick velocity, 
$<V_k^2>^{1/2}$, for both sub- and super-Eddington systems (filled circles)  
and for sub-Eddington systems only (open circles). For smaller common-envelope 
efficiencies, $\alpha_{CE}$, the ratios are higher.} 

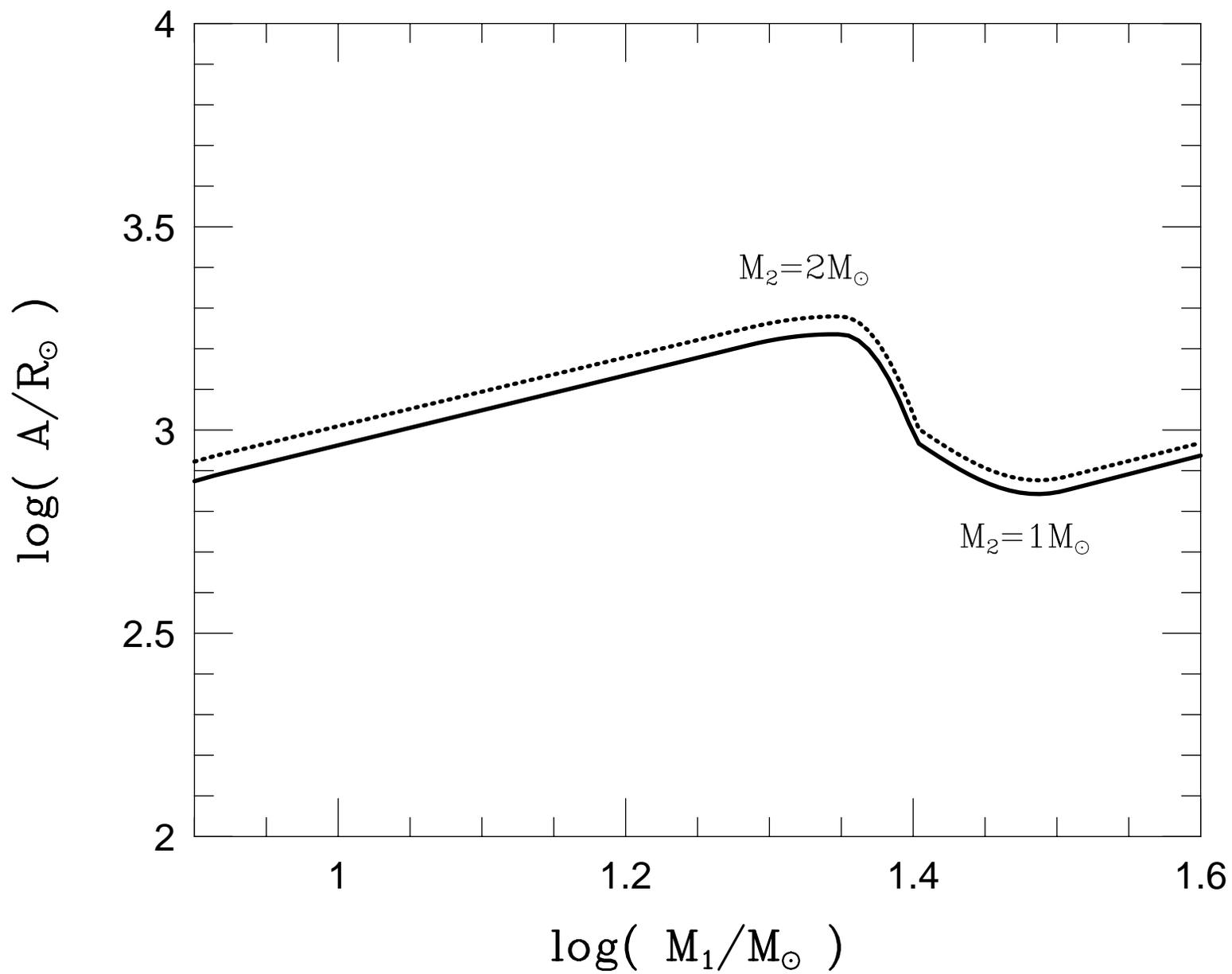
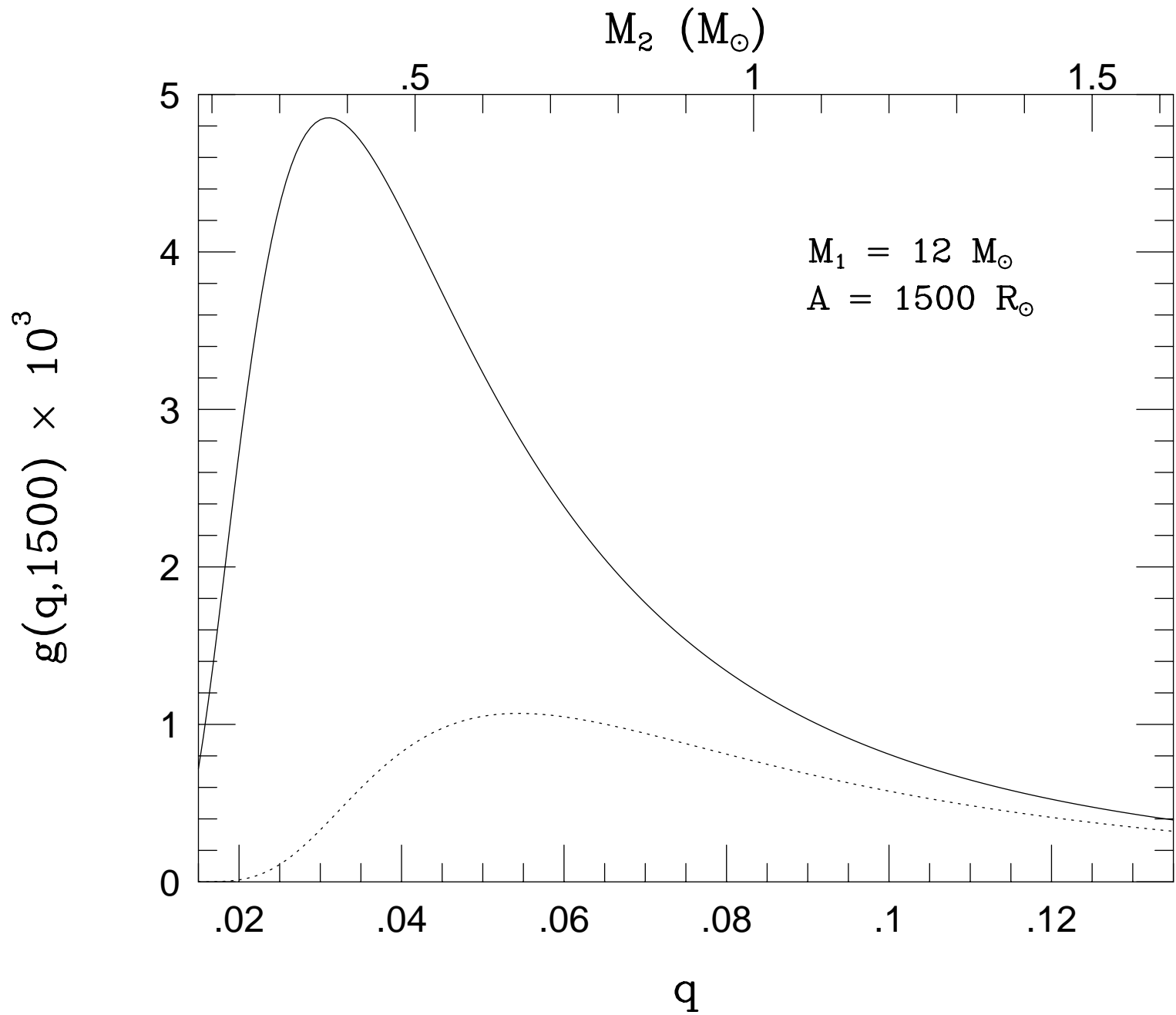
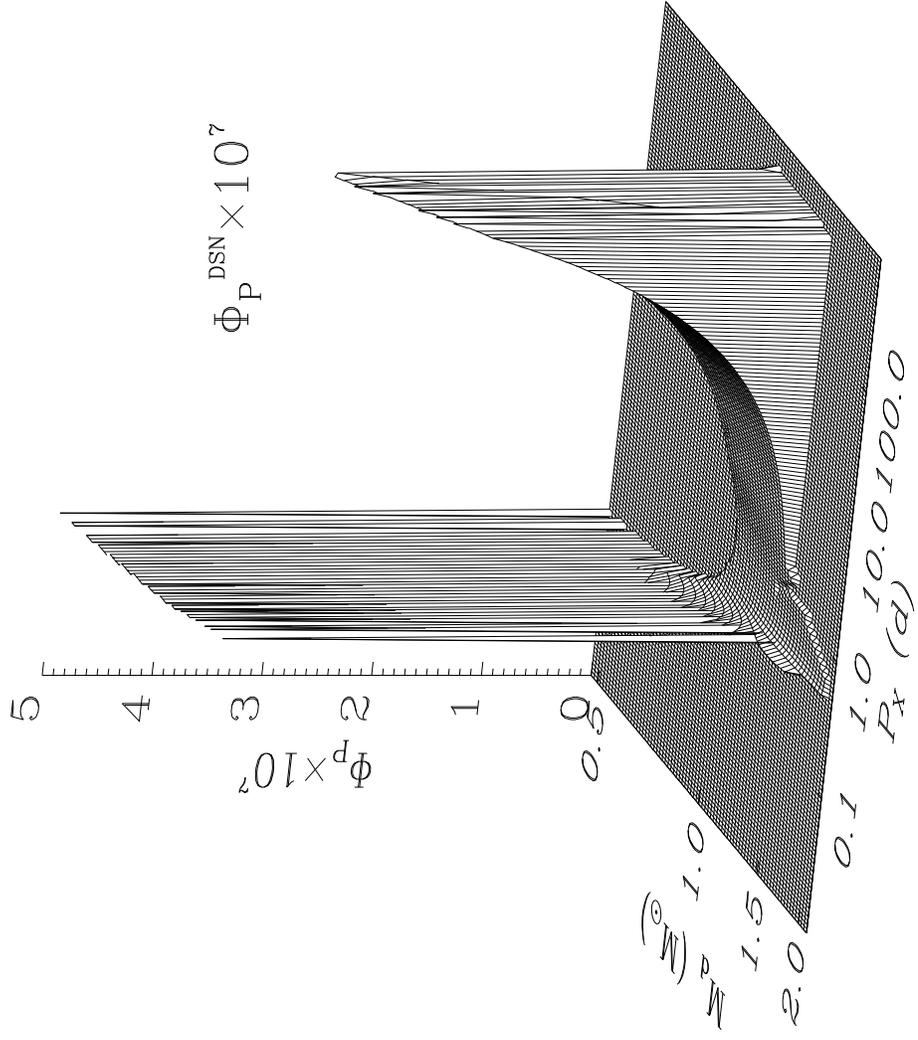
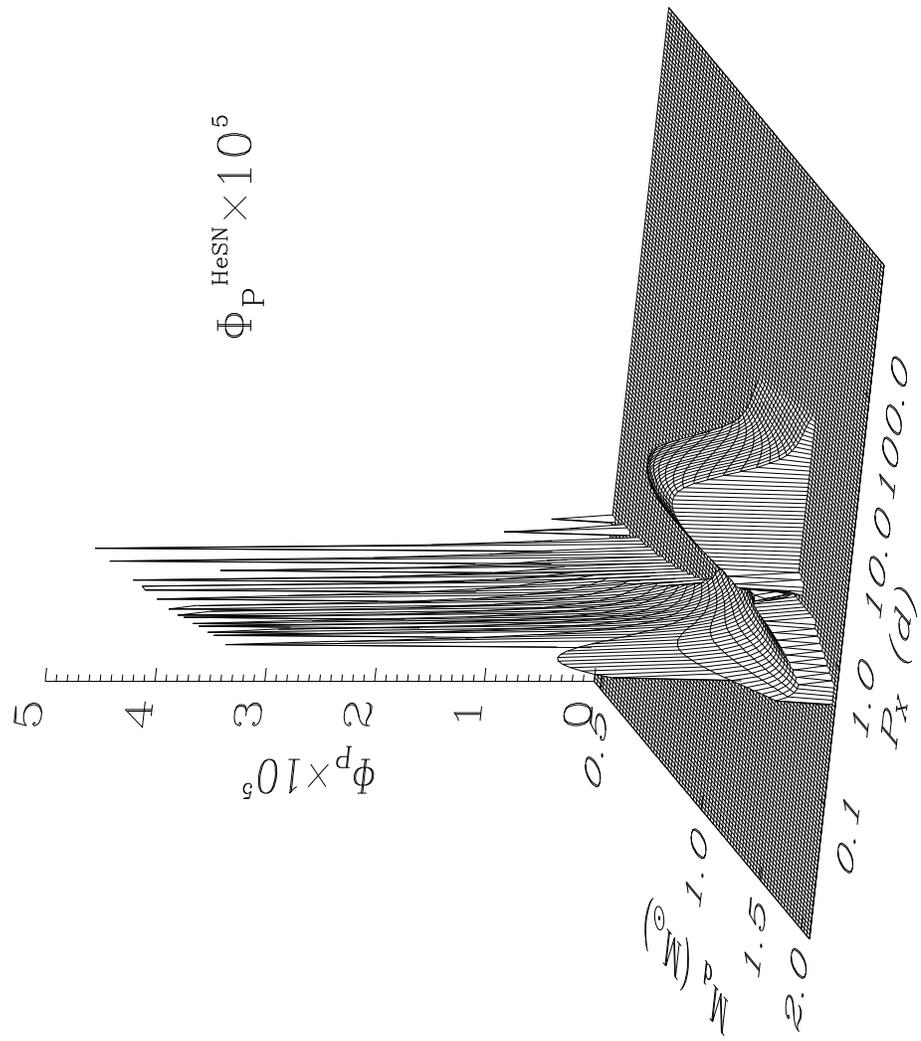
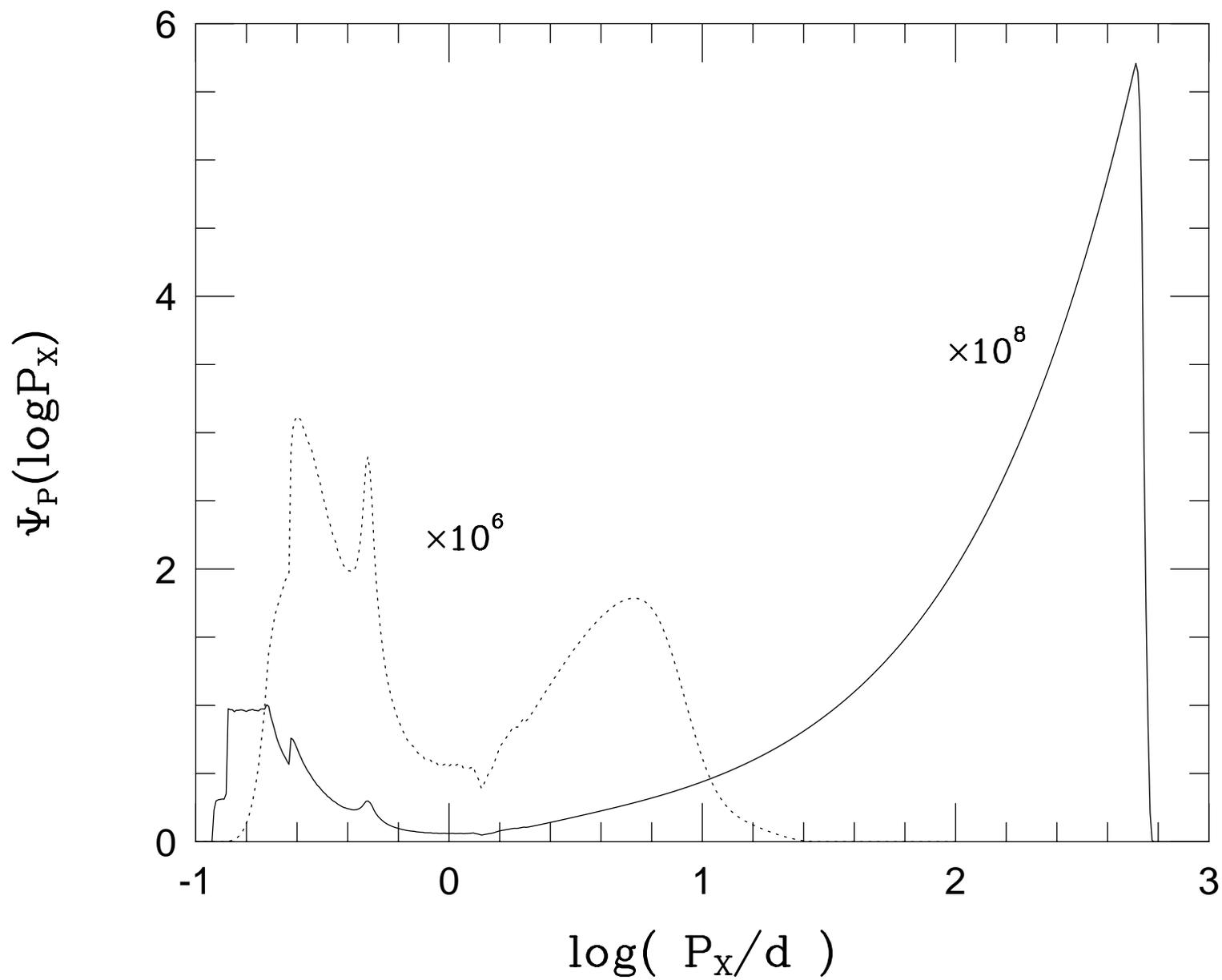
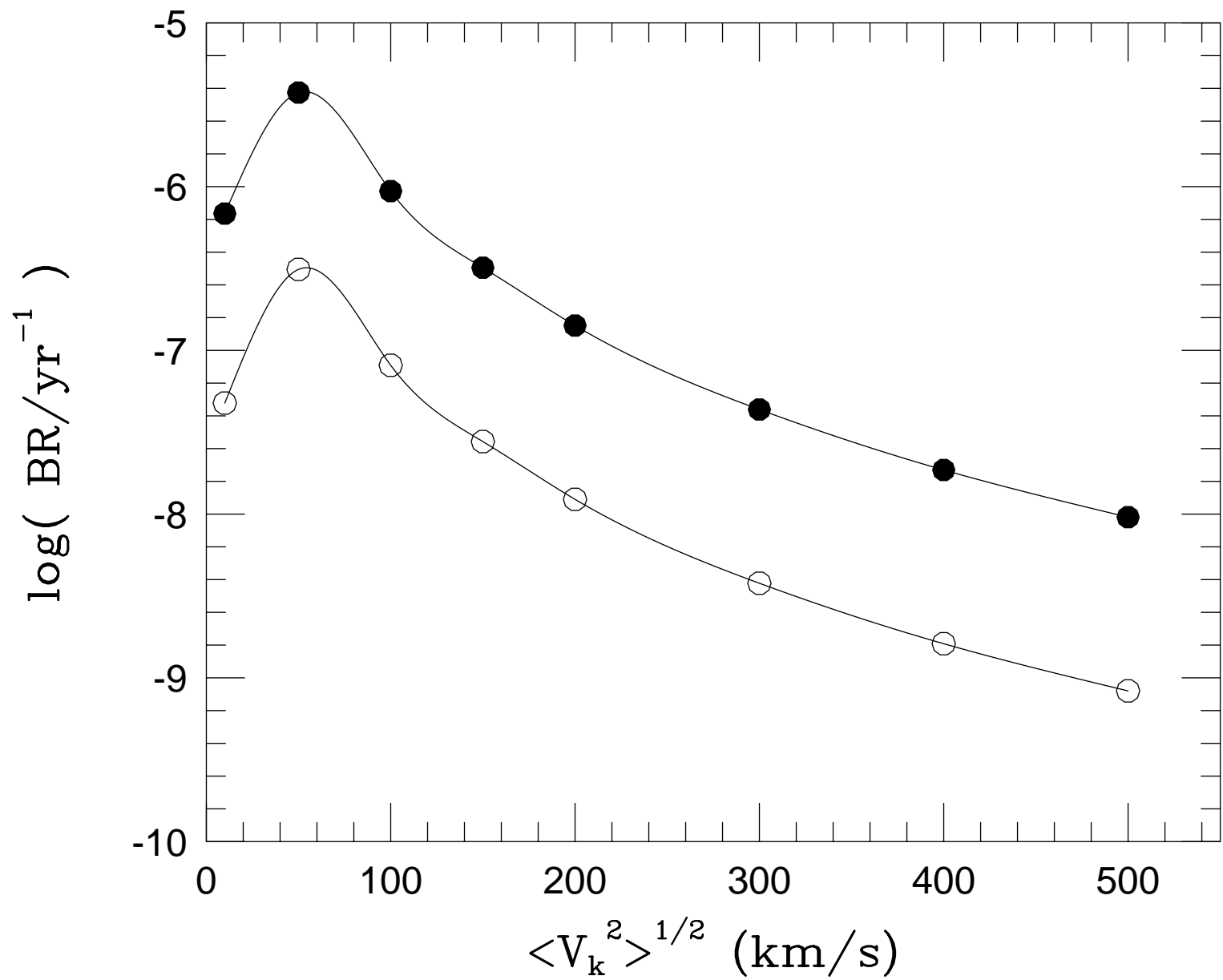
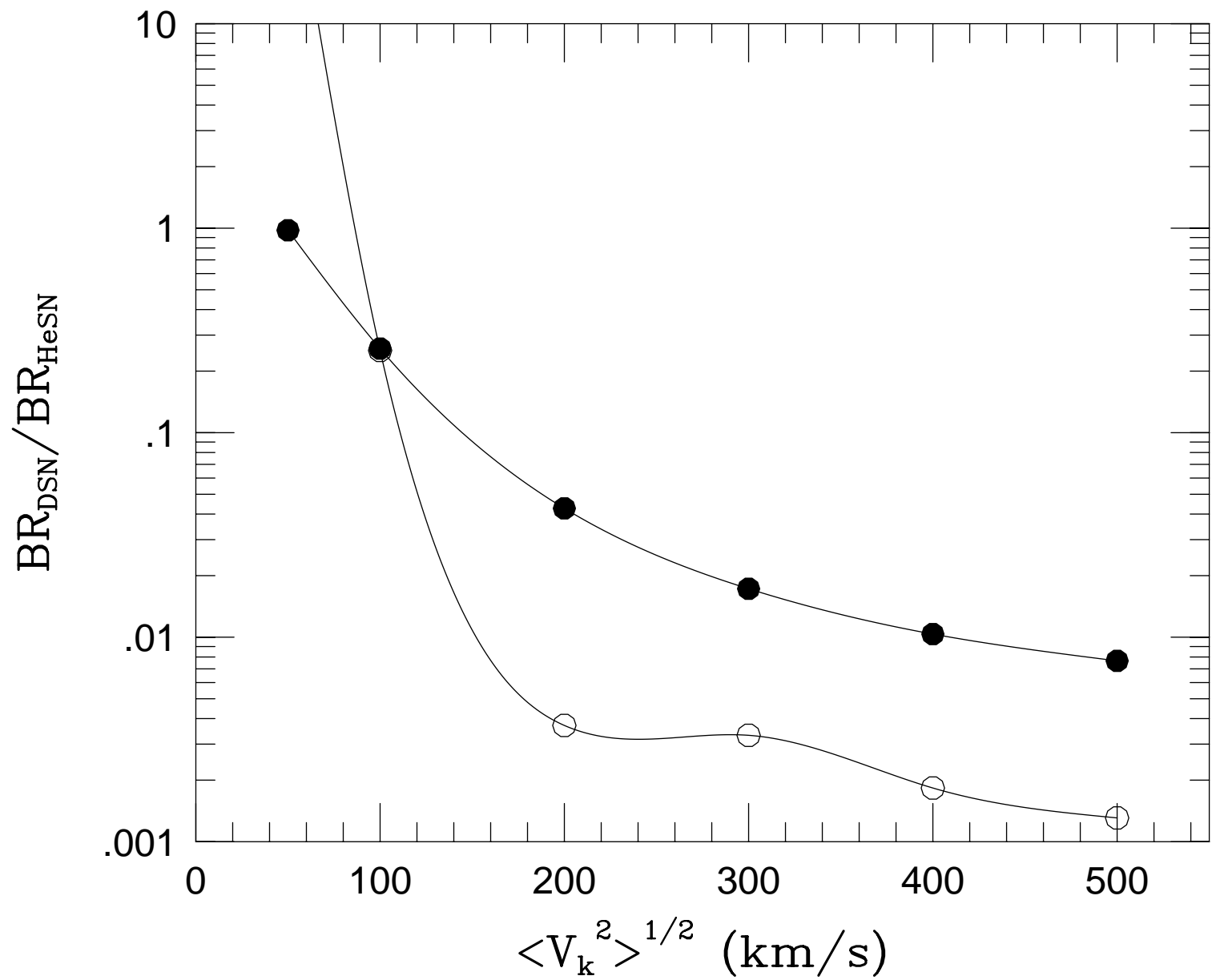
\end{document}